\documentclass[aps,prl,twocolumn,showpacs,groupedaddress]{revtex4}
\usepackage{graphicx}

\begin{document}

\title{Geometric  phase in   dephasing systems}
\author{X. X. Yi, L. C. Wang, and W. Wang}
\affiliation{Department of physics, Dalian University of
Technology, Dalian 116024, China}

\date{\today}

\begin{abstract}
Beyond the quantum Markov approximation, we calculate the
geometric phase of a two-level system driven by a quantized
magnetic field subject to phase dephasing. The  phase reduces to
the standard geometric phase in the weak coupling limit and it
involves the phase information of the environment  in general. In
contrast with the geometric phase in dissipative systems, the
geometric phase acquired by the system can be observed on a long
time scale. We also show that with the system decohering to its
pointer states, the geometric phase factor tends to a sum over the
phase factors pertaining to the pointer states.
\end{abstract}

\pacs{ 03.65.Vf, 03.65.Yz} \maketitle

Quantum mechanics tell us that physical states are equivalent up
to a global phase, which in general does not contain useful
information about the described system and thus can be ignored.
This is not the case, however, for a system transported round a
circuit by varying the parameters $\vec{s}=(s_1, s_2,...)$ in its
Hamiltonian $H(\vec{s})$. As Berry showed \cite{berry84}, the
phase can have a component of geometric origin called geometric
phase with important observable consequences, such as the
Aharonov-Bohm effect \cite{aharonov59} and the spin-$\frac 1 2 $
particle driven by a rotating magnetic field \cite{berry84}. The
geometric phases that only depend on the path followed by the
system during its evolution, have been investigated and tested in
a variety of settings and have been generalized in several
directions \cite{shapere89}. The geometric phases are attractive
both from a theoretical perspective, and from the point of view of
possible applications, among which geometric quantum computation
\cite{zanardi99,jones99,ekert00,falci00} is one of the most
importance.

As realistic systems always interact with their environment, the
study on the geometric phase in open systems become interesting.
Garrison and Wright \cite{garrison88} were the first to touch on
this issue by describing open system evolution in terms of a
non-Hermitian Hamiltonian. This is a pure state analysis, so it
did not address the problem of geometric phases for mixed states.
Toward the geometric phase for mixed states in open systems, the
approaches used involve solving the master equation of the system
\cite{fonsera02,aguiera03,ellinas89,gamliel89,kamleitner04},
employing a quantum trajectory analysis \cite{nazir02,carollo03}
or Krauss operators \cite{marzlin04}, and the perturbative
expansions \cite{whitney03,gaitan98}. Some interesting results
were achieved, briefly summarized as follows: nonhermitian
Hamiltonian lead to a modification of Berry's phase
\cite{garrison88, whitney03}, stochastically evolving magnetic
fields produce both energy shift and broadening \cite{gaitan98},
phenomenological weakly dissipative Liouvillians alter Berry's
phase by introducing an imaginary correction \cite{ellinas89} or
lead to  damping and mixing of the density matrix elements
\cite{gamliel89}. However, almost all these studies are performed
for dissipative systems, and thus the representations are
applicable for systems whose energy is not conserved. For open
systems with conserved energy (dephasing systems), the problem
beyond the Markov approximation  remains untouched  to our best
knowledge. Because the system-environment interaction $H_I$ and
the free system Hamiltonian $H_s$ commute for dephasing systems,
the dynamical problem and then the geometric phase of the system
may be solved/calculated precisely.  On the other hand, the
previous study \cite{whitney03} shows that one can not perform an
arbitrarily long experiment to measure the phase for dissipative
systems, i.e., it is not allowed  to draw phase information out of
the system on a long time scale, this feature of dissipative
systems again motivates  investigation on the problem of geometric
phases in dephasing systems, where in principle one may get
analytical results for the phase on any time scale.

In this Letter, we investigate the behavior of the geometric phase
of a two-level system interacting with a driving magnetic field
when this system is not only quantized but also subjected  to
decoherence. The environment that leads to decoherence may
originate from the fluctuation in the driving fields, or from the
vacuum fluctuations, or from the background radiations. We
calculate and analyze the effect of dephasing of the driven system
on the geometric phase of the system, our discussions  will
distinguish between two kinds of evolution: (1)The environment
undergoes an adiabatic evolution, and (2)it evolves as it may.

Let us consider a two-level system driven by a quantized magnetic
field and subjected to decoherence. The decoherence process is
described by the coupling of the two-level system to an
environment of harmonic oscillators with frequencies $\{ \omega_j
\}$, the Hamiltonian governing such a system reads \cite{bose98}
\begin{eqnarray}
H&=&H_s+H_I+H_e,\nonumber\\
H_s&=&\hbar\frac{\Omega}{2}\sigma_z+\hbar\Omega
a^{\dagger}a+\hbar g(\sigma_+ a+\sigma_-a^{\dagger}),\nonumber\\
H_I&=&(\sigma_+
a+\sigma_-a^{\dagger})\sum_j\hbar\lambda_j(b_j^{\dagger}+b_j),\nonumber\\
H_e&=&\hbar\sum_j\omega_jb_j^{\dagger}b_j, \label{ha1}
\end{eqnarray}
where $a$, $a^{\dagger}$ are boson operators for the driving
field, $\sigma_+$, $\sigma_-$, and $\sigma_z$ are pauli operators
for two relevant internal atomic levels $|e\rangle$ and
$|g\rangle$, and $b_j^{\dagger}$, $b_j$ are the creation and
annihilation operators of the environment bosons. The system
hamiltonian $H_s$ characterizes Jaynes-Cummings dynamics without
neither energy relaxation nor phase dephasing, while terms $H_e$
and $H_I$ describe a bosonic environment and its coupling to the
Jaynes-Cummings system. The choice of the coupling between the
system   and the environment   determines   effects of the
environment. For example, the choice of the system operators that
do not change the quantum number of the driving field when they
operate on the dressed state would result in relaxations within
the dressed states indicated by the quantum number $n$, but not
energy relaxations between states with different $n$. The
system-environment coupling chosen in our model is exactly such a
choice, and thus, we may rewrite the Hamiltonian Eq. (\ref{ha1})
in terms of the dressed states as
\begin{eqnarray}
H&=&\bigoplus_n H_n,\nonumber\\
H_n&=&E_+(n)|+(n)\rangle\langle
+(n)|\nonumber\\
&+&E_-(n)|-(n)\rangle\langle
-(n)|+\sum_j\hbar\omega_jb_j^{\dagger}b_j\nonumber\\
&+&\sqrt{n+1}(|+(n)\rangle\langle
+(n)|\nonumber\\
&-&|-(n)\rangle\langle
-(n)|)\sum_j\hbar\lambda_j(b_j^{\dagger}+b_j),
\end{eqnarray}
where $E_{\pm}(n)=\frac{2n+1}{2}\hbar\Omega\pm \hbar g\sqrt{n+1}$,
and the dressed states with indication $n$ are
\begin{equation}
|\pm(n)\rangle=\frac{1}{\sqrt{2}}(|g,n+1\rangle\pm |e,n\rangle).
\end{equation}
Here $\{|n\rangle\}$ stand for the Fock states of the driving
field. The Hamiltonian $H_n$ describes a driven harmonic
oscillator,  the driving terms depend on the dressed states, but
they are independent of the dressed state indication $n$. With
these properties, we may simplify the problem and restrict  our
study within the dressed states with the same indication $n$. The
eigenstates of the Hamiltonian $H_n$ take the following form,
\begin{eqnarray}
|e_+(n)\rangle_{\{n_j\}}&=&|+(n)\rangle\otimes
\prod_j|n_j\rangle_{B_{+,j}},\nonumber\\
|e_-(n)\rangle_{\{n_j\}}&=&|-(n)\rangle\otimes
\prod_j|n_j\rangle_{B_{-,j}},\label{eist}
\end{eqnarray}
where $|n_j\rangle_{B_{\pm,j}}$ denotes the Fock state for new
environment mode $B_{\pm,j}=(b_j \pm
\lambda_j\sqrt{n+1}/\omega_j)|\pm(n)\rangle\langle\pm(n)|,$ i.e.,
$B_{\pm,j}^{\dagger}B_{\pm,
j}|n_j\rangle_{B_{\pm,j}}=n_j|n_j\rangle_{B_{\pm,j}}$, depending
on the dressed states. The corresponding eigenenergies are $\hbar
g\sqrt{n+1}+\sum_j\hbar \omega_j n_j$ and
 $-\hbar g\sqrt{n+1}+\sum_j\hbar \omega_j n_j$, respectively.
We distinguish between two situations to study the geometric phase
of the system, the first is to consider the universe
(system+environment) to undergo an adiabatic evolution, in this
case the states of the environment never evolve during the
evolution. Because the relaxation of the system due to its
coupling to the environment is independent of the dressed state
indication $n$,   the acquired geometric phase of the open system
are the same as in the case without  the environment. The second
situation is more practical, in which we assume the environment
initially is in its ground state $
\prod_j|0_j\rangle_{b_j}=\prod_j|\pm\frac{\lambda_j\sqrt{n+1}}{\omega_j}\rangle_{B_{\pm,j}}^c$
and evolve   govern by the Hamiltonian Eq. (\ref{ha1}). This
initial state means that the environment is initialized in the
vacuum of modes $\{b_j\}$, or in coherent state
$|\pm\frac{\lambda_j\sqrt{n+1}}{\omega_j}\rangle_{B_{\pm,j}}^c$ of
modes $\{B_{\pm,j}\}$. The geometric phase of the universe in this
situation may be calculated by removing the accumulation of these
dynamical phase from the total phase, i.e.,
\begin{equation}
\phi_g(n)=\mbox{arg}\langle\Psi(0)|\Psi(T)\rangle+i\int_0^T
dt\langle \Psi(t)|\frac{\partial}{\partial t}|\Psi(t)\rangle,
\label{bp}
\end{equation}
it is easy to demonstrated that for closed systems $\phi_g(n)$
reduces to the Aharonov-Anandan formula for cyclic evolutions
\cite{aharonov87} and to the Berry phase for adiabatic and cyclic
evolutions \cite{berry84}. The initial state together with the
time evolution operator determine the path followed by the system,
in this sense the geometric phase might depend on the initial
condition. If we choose $|\Psi(0)\rangle=|+(n)\rangle\otimes
\prod_j|\frac{\Lambda_j}{\omega_j}\rangle_{B_{+,j}}^c$
($\Lambda_j=\lambda_j\sqrt{n+1}$) as the initial state, at time
$t$ the universe   evolves to
\begin{equation}
|\Psi(t)\rangle=|+(n)\rangle\otimes
\prod_j|\frac{\Lambda_j}{\omega_j}e^{-i\omega_jt}\rangle_{B_{+,j}}^c.
\end{equation}
In contrast with classically driving field, in this study the
driving field is quantized. In order to generate a phase change in
the state of the field, we borrow the idea in Ref.\cite{fuentes02}
to introduce the phase shift operator $U(\psi)=exp(-i\psi
a^{\dagger}a)$ and adiabatically apply it to the Hamiltonian of
the system. Changing $\psi=\bar{\Omega} \cdot t$ slowly from $0$
to $2\pi$ (the corresponding time from $0$ to
$T=\frac{2\pi}{\bar{\Omega}}$) the geometric phase generated is
calculated by Eq. (\ref{bp}) as follows,
\begin{eqnarray}
\phi_g^+(n)&=&\mbox{arg}[\prod_je^{-|\frac{\Lambda_j}{\omega_j}|^2(1-e^{-i\omega_j
T})}]\nonumber\\
&+&\sum_j[ \omega_j T
e^{-|\frac{\Lambda_j}{\omega_j}|^2}\sum_{m=0}^\infty
m\frac{(\Lambda_j/\omega_j)^{2m}}{m!}]\nonumber\\
&+&\gamma_+(n), \label{gp1}
\end{eqnarray}
where $\gamma_+(n)=(2n+1)\pi$ is the Berry phase acquired when the
universe remains unchanged. For a continuous spectrum of the
environmental modes, the sum over $j$ in the above expressions is
replaced by an   integral involving the spectral density with  a
cutoff frequency $\omega_c$
\begin{equation}
\rho(\omega)=\varepsilon (\frac{\omega}{\lambda_{\omega}})^2,
0\leq \omega \leq \omega_c, \label{spe}
\end{equation}
this spectrum density is of Ohmic type. Eq. (\ref{bp}) and
Eq.(\ref{spe}) together yield
\begin{equation}
\phi_g^+(n)=\gamma_+(n)+ \varepsilon\frac{\omega_c^2(n+1)T}{2}+
\frac{(n+1)}{T}\varepsilon[\cos(\omega_c T)-1].\label{gp2}
\end{equation}
If we choose the eigenstate $|\Psi(0)\rangle=|-(n)\rangle\otimes
\prod_j|-\frac{\Lambda_j}{\omega_j}\rangle_{B_{-,j}}^c$ from
another set of eigenstates Eq. (\ref{eist}) as the initial
condition, the time evolution of the universe can be expressed as,
\begin{equation}
|\Phi(t)\rangle=|-(n)\rangle\otimes
\prod_j|-\frac{\Lambda_j}{\omega_j}e^{-i\omega_jt}\rangle_{B_{-,j}}^c.
\end{equation}
In the same way, we can get the geometric phase pertaining to this
evolution loop,
\begin{eqnarray}
\phi_g^{-}(n)=\phi_g^+(n)-\gamma_+(n)+\gamma_-(n),\label{gp3}
\end{eqnarray}
namely, the contributions from the system-environment coupling are
the same for the both pathes.  Here $\gamma_-(n)$ is the Berry
phase attaining to the dressed state $|-(n)\rangle$ in the case
without environment, it is easy to prove that it takes the same
expression as $\gamma_+(n)$.

The last two terms in   equation Eq. (\ref{gp2})  result from the
system-environment couplings, they vanish when the couplings tend
to zero (in the equations, $\omega_c\rightarrow 0$), thus the
  expressions return to the geometric phases presented in
Ref. \cite{fuentes02}.  The path the system followed changes the
system-enviroment coupling, and the environment is impossible to
return to its initial state due to its huge variety of freedom,
the geometric phase then depends on the time $T$ when we draw out
phase information from the system. In the classical limit $n
\rightarrow \infty$, the contributions from the system-environment
coupling tend to infinity  caused by relative strong coupling
between the system and the driving field, hence it becomes
undefined in the sense of interferometry.  It is interesting to
note that for $n=0$ the phases are not zero, which means that the
vacuum driving field introduces a correction in geometric phases,
this expression is relevant when systems are driven by fields with
few photons. For weak system-environment coupling \cite{note1},
Eq. (\ref{gp1})( Eq. (\ref{gp3}), in the same way) may be expanded
in powers of $\Lambda_j/\omega_j$, up to the second order of
$\Lambda_j/\omega_j$, Eq.(\ref{gp1}) follows,
\begin{equation}
\phi_g(n)=\gamma_+(n)+\sum_j \omega_j T
(\frac{\Lambda_j}{\omega_j})^2.
\end{equation}
The explanation of this result is very simple, as the
system-environment coupling is very weak, the most contribution to
the geometric phase come from the system-driving field coupling.
The same dependence of $\phi_g(n)$ on $\Lambda_j$ (i.e.,
$\Lambda_j^2$, no contribution proportional to $\Lambda_j$) can be
found in Ref. \cite{whitney03}.

Up to now, we have calculated the geometric phase for the
universe, explicit expressions for the geometric phase were
obtained, these expressions are of relevance for the case in which
systems are coupled to an environment that describes parameter
fluctuations. For instance, imperfect dipole transitions between
states $|g\rangle$ and $|e\rangle$ due to fluctuation of the
driving laser intensity may be modelled by the coupling in Eq.
(\ref{ha1}), in this situation the environment and the driving
field are the same. This is not the case, however,  when the
environment is an independent system (say, black body radiations),
we have to trace out the environment in order to calculate the
dynamical information for the system, thus the evolution of the
system is no longer unitary. For non-unitary evolution, the
geometric phase can be calculated as follows. First, solve the
eigenvalue problem for the reduced density matrix $\rho(t)$ and
obtain its eigenvalues $\varepsilon_k(t)$ as well as the
corresponding eigenvectors $|\psi_k(t)\rangle;$ secondly,
substitute $\varepsilon_k(t)$ and $|\psi_k(t)\rangle$ into
\begin{widetext}
\begin{equation}
\Phi_g(n)=\mbox{arg}(\sum_k\sqrt{\varepsilon_k(0)\varepsilon_k(T)}\langle\psi_k(0)|\psi_k(T)\rangle
e^{-\int_0^T\langle\psi_k(t)|\partial/\partial t|\psi_k(t)\rangle
dt}).\label{gp5}
\end{equation}
\end{widetext}
Here, $\Phi_g(n)$ is the geometric phase for the system undergoing
non-unitary evolution \cite{tong04}. The geometric phase Eq.
(\ref{gp5}) is gauge invariant  and can be reduced to the
well-known results in the unitary evolution, thus it is
experimentally testable. Now, we exploit the expression to
calculate the geometric phase for the driven two-level system.  To
this aim, we first write down the reduced density matrix
$\rho(t)$,
\begin{equation}
\rho(t)=\left( \matrix{ |c_+|^2 & c_+^*c_-F(t)  \cr c_+c_-^*F^*(t)
& |c_-|^2 \cr } \right),
\end{equation}
where the initial state of
$(c_+|+(n)\rangle+c_-|-(n)\rangle)\otimes\prod_j
|0_j\rangle_{b_j}$ is assumed for the universe, and
$F(t)=exp[-\frac{i}{\hbar}(E_+(n)-E_-(n))t]\cdot
exp[-\sum_j\eta_j(t)]$ with
$\eta_j(t)=4|\frac{\Lambda_j}{\hbar\omega_j}|^2(1-\cos\omega_j
t)$. Simple algebra gives the eigenvalues and the corresponding
eigenvectors of $\rho(t)$,
\begin{eqnarray}
\varepsilon_{\pm}(t)&=&\frac 1 2 \pm \frac 1 2
\sqrt{(|c_+|^2-|c_-|^2)^2+4|c_+^*c_- F(t)|^2},\nonumber\\
|\psi_{\pm}(t)\rangle&=&X_{\pm}(t)|+(n)\rangle+Y_{\pm}(t)|-(n)\rangle,\label{eifv}
\end{eqnarray}
where
$X_{\pm}(t)=c_+^*c_-F(t)/\sqrt{|c_+^*c_-F(t)|^2+(\varepsilon_{\pm}(t)-|c_+|^2)},$
and $Y_{\pm}(t)=\sqrt{1-|X_{\pm}(t)|^2}$. Eq. (\ref{eifv}) and
Eq.(\ref{gp5}) together yield the geometric phase for the system.
The expression for the geometric phase is tedious, so instead of
writing down the expression, we present here some remarkable
comments. The expressions for the geometric phase are analytically
exact, so we might predict the behavior of the geometric phase on
a long time scale. For any $j$, $\eta_k(t)\geq 0$, so with
$t\rightarrow \infty$, $F(t)\rightarrow 0$ for a random spectrum
density of the environment, this indicates that
$\Phi_g(T\rightarrow\infty)\rightarrow
\mbox{arg}[|c_+|^2e^{-i\gamma_+(n)}+|c_-|^2e^{-i\gamma_-(n)}]$ for
a relative long $T$. This result is quite interesting: With the
off-diagonal elements of the reduced density matrix tending to
zero (decoherence), the driven two-level system  decohers to its
pointer states $|+(n)\rangle$ or $|-(n)\rangle$, while the
geometric phase factor of the driven system  reduces to a weighted
sum over the phase factors pertaining to the pointer states, this
provides us a new way to observe decoherence effects.

Summarizing, the geometric phases for a dephasing system (open
system) have presented and discussed. The open system has been
demonstrated by a driven two-level system  coupling to an
environment of harmonic oscillators. The results show that there
are no correction in the geometric phase of the universe due to
the system-environment coupling when the environment undergoes an
adiabatic evolution, whereas the correction in the phase is path
dependent when the constraint on the  evolution  is released. The
mixed state geometric phase for the dephasing system is also
presented and discussed. The geometric phase factor would tend to
a sum over these phase factor pertaining to the pointer states
with the open system decohering  to its pointer states, it is a
reflection of the decoherence in geometric phases.
\ \ \\
Comments from Dr. Peter Marzlin and simulating discussions with
Dr. Robert Whitney  are gratefully acknowledged. This work was
supported by EYTP of M.O.E, and NSF of China Project No. 10305002.\\

\end{document}